# A New Rapid Test to End COVID

Baker, M.R.[a], Conway, F.[a], Dal Ben, F.[a], Hawthorne, E.L.[a], Iacoviello, L.[b,c],
Agodi, Antonella.[d], Mukhtar, Saqib.[a], Phan Hai.H.[a], Rabvukwa, Y.[a]., Rogge, J.R.[a]

a Complex Systems Research Division, Medichain Ltd. Euston London, UK
b Department of Epidemiology and Prevention, IRCCS NEUROMED, Pozzilli (IS), Italy.
c Research Center in Epidemiology and Preventive Medicine (EPIMED). Department of Medicine and Surgery.
University of Insubria, Varese, Italy.
d Department of Medical and Surgical Sciences and Advanced Technologies "G.F. Ingrassia", University of Catania;
AOU Policlinico-Vittorio Emanuele, Catania. Italy

Despite 93.1% to 95.8% of the UK adult population having been vaccinated and currently 83.5% to 89.8% of adults having received at least two doses (1), and despite many households testing twice a week with lateral flow tests (2), R at the time of writing is 0.9 to 1.1, with a growth rate range for England of between -1% and +1% (3). Furthermore, up to 30% of infected individuals are going on to experience Long Covid (4). The crisis is far from over and as new variants of concern like Omicron spread, the situation is not under control, even in the highly vaccinated and tested UK and far less so in many countries. The problem is likely to be replicated in other countries with currently low infection levels as isolation is eased in future, even if these countries reach a high level of vaccination. Additionally, concerns have been raised about fall in immunity by 6 months after receiving the Pfizer and AstraZeneca vaccines (5), and the ability of Omicron to re-infect and cause illness in vaccinated people, with urgent booster jabs now being given to attempt to mitigate this. A solution to stop the spread of all variants of COVID-19 is needed now, and we present it here: CLDC, a rapid test that is 98%+ sensitive, low cost and scalable.

The intention is to identify SARS-CoV-2 infection more reliably than either lateral flow or PCR (98%+ sensitivity vs lateral flow and PCR sensitivity of only 40-70% in real world testing (6,7)).

CLDC testing should be used on a daily basis by those wishing to leave their homes, with strict self isolation for those testing positive. This frequent, reliable mass testing is the only viable way to remove sufficient numbers of infectives from circulation to eradicate the virus - before further mutations lead to vaccine escaping variants, rising hospitalisations and an ongoing cycle of lockdowns and uncertainty.

A crucial combination of factors distinguish CLDC from other tests, which fail to identify enough cases to stop the spread without community-wide restrictions: its ability to test everyone daily due to its low cost and high scalability, its immediate result, and its higher sensitivity, including its potential to identify the presence of the virus earlier in the course of infection, when carriers may be infectious but are pre-symptomatic and test negative on other tests such as lateral flow and PCR.

In this study we have applied a variation of the previously reported CLDC (Characteristic Leukocyte Differential Count) algorithmic ensemble to form the basis for the detection of SARS-CoV-2 via a high sensitivity, low-cost rapid test based on a specific early immune response to SARS-CoV-2 expressed in the differential white blood count (8). In this version decreasing levels of Lymphocytes, Eosinophils, and increases in Neutrophil levels, are heuristically combined in an ensemble algorithm which also included L-N (Leucocyte minus Neutrophil), and (Leucocyte minus Eosinophil) test stages.

Using training data from the COvid-19 RISk and Treatments (9) collaboration and the Moli-Sani Project, both from southern Italy, comprising WBC differential data for 72 positive (RT-PCR confirmed) and 4742 negative COVID patients, respectively, two variations of the test were applied. The first (algorithmic ensemble CLDC 21-04-SE) was optimised for sensitivity and has a sensitivity of 97·29% and negative percent agreement of 67·95% for a single test (specificity estimated at 70.4%). This performance is similar to that of the originally published algorithm produced with data from the Albert Einstein Hospital in São Paulo, which had a sensitivity of 98·7% and negative percent agreement of 63·8% (8). A version of the algorithmic ensemble optimised to reduce false-positive rate (algorithmic ensemble CLDC 21-04-SP) has a sensitivity of 81·82% with a specificity of 96·50%. Thus this overcomes the widely reported issues of lateral flow tests (10) missing large proportions of positive cases (giving false negative results), while also exceeding the specificity of 82% (18% false positive results) from 30,904 positive results in 26m lateral flow tests recorded by Public Health England in March 2021 (11,12).

Our CLDC algorithms were also applied to data from the Regional Service for the Epidemiology and Surveillance of Infectious Diseases database, a longitudinal cohort study of 379 patients with laboratory-confirmed COVID-19 admitted at the Italian National Institute for Infectious Diseases "Lazzaro Spallanzani" (INMI), in Rome (Italy) (13). The 21-04-SE algorithm showed a sensitivity of 98·64% during the first seven days during symptom onset and 96·73% sensitivity for data taken over 21 days after initial symptoms.

Suggested methodological improvements from the Centre for Mathematical Modelling of Infectious Diseases at the London School of Hygiene & Tropical Medicine indicate that it could be possible to improve the false positive rate of our 21-04-SP test to 1.87% (14) in the field.

Our rapid test, which results from this methodology and which has been field trialed at the Chelsea Pharmacy Medical Clinic, London, can return these results in less than five minutes using a point-of-care five-part differential blood count system and our analysis portal, preventing delays in isolating cases and their contacts due to RT-PCR test delays, thus reducing transmission risk.

These studies indicate that the CLDC 21-04-SE and CLDC 21-04-SP algorithms form the basis for a test that can detect SARS-CoV-2 with a higher level of sensitivity and specificity than existing testing (15), and have the ability to selectively trade off sensitivity for specificity by adjusting algorithm ensemble parameters, based on a well-established test process which can be significantly cheaper and faster than existing tests and control the issues surrounding false positives in current rapid testing (16).

Furthermore, evidence indicates that the predictive value of the current gold standard RT-PCR testing varies with time from exposure and symptom onset in a way that falsely reassures by giving negative test results which may not be valid (17). In Kucirka et al's study the probability of a false-negative RT-PCR result in an infected person was shown to be 100% (95% CI, 100% to 100%) four days before symptom onset, 67% (CI, 27% to 94%) one day before and 38% (CI, 18% to 65%) on the day of symptom onset (17). Compared with this, early indications suggest that CLDC may have the potential for improved early detection owing to the changes in leucocyte ratios in the early stages of infection (18). Modelling shows that the sensitivity and specificity advantages over other rapid diagnostic tests give a substantial reduction in R, which would be of substantial value in decreasing viral transmission. Following that, our self-administered home version of the test would allow the scale and reliability required to eliminate the virus.

### Declaration of competing interests

MB, FC, FD, EH, SM, HB, YR are supported by Innovate UK Grant 73420. MB is a named inventor on two UK patents.

### Acknowledgements

Funding: Supported by an Innovate UK Grant 73420. Innovate UK is part of UK Research and Innovation (UKRI).